\documentclass[11pt]{article}
\usepackage{latexsym}
\usepackage{graphicx}
\usepackage{amssymb}
\usepackage{amsmath,enumerate,rotate}

\def\R{{\rm{I}\! \rm{R}}}

\def\pRR{\hbox{{\tiny \rm I}\kern-.1em\hbox{{\tiny \rm R}}}}

\def\NN{\hbox{I\kern-.2em\hbox{N}}}

\newcommand{\mR}{\mathcal R}
\newcommand{\mN}{\mathcal N}

\newcommand{\mT}{\mathcal T}
\newcommand{\mI}{\mathcal I}

\newcommand{\fer}[1]{(\ref{#1})}
\newcommand{\be}{\begin{equation}}
\newcommand{\ee}{\end{equation}}
\def\bqa{\begin{eqnarray}}
\def\eqa{\end{eqnarray}}
\def\bd{\begin{displaymath}}
\def\ed{\end{displaymath}}

\newtheorem{theorem}{\bf Theorem}[section]

\begin{document}

\title{R\'enyi entropies and nonlinear diffusion equations}
\author{ G. Toscani\thanks{Department of Mathematics,
University of Pavia, via Ferrata 1, 27100 Pavia, ITALY.} }

\maketitle

\begin{abstract}
Since their introduction in the early sixties \cite{Ren},  the R\'enyi entropies have been used in many contexts, ranging from information theory to astrophysics, turbulence phenomena and others.
In this note, we enlighten the main connections between R\'enyi entropies and nonlinear diffusion equations. In particular, it is shown that these relationships allow to prove various functional inequalities in sharp form.


\end{abstract}

\section{Introduction}
\label{intro}
Given a probability density $f(x)$, $x \in \R^n$, and a positive constant $p$
the R\'enyi entropy of order $p$ of  $f$  is defined by \cite{DCT}:
 \be\label{re}
\mR_p(f) = \frac 1{1-p} \log\left( \int_{\pRR^n} f^p(y) \, dy \right).
 \ee
This concept of entropy has been introduced by R\'enyi in \cite{Ren} for a discrete pro\-ba\-bility measure to generalize the classical logarithmic entropy, by maintaining at the same time most of its properties. Indeed, the R\'enyi entropy of order $1$,  defined as the limit as $p \to 1$ of
$\mR_p(f)$ is
 \be\label{shan}
 \lim_{p\to 1} \mR_p(f) = \mR(u) = -\int_{\pRR^n}f(y) \log f(y) \, dy .
 \ee
Therefore, the standard (Shannon) entropy  of a probability density \cite{Sha} is included in the set of R\'enyi entropies, and it is identified with the R\'enyi entropy of index $p = 1$.

Among other properties, the R\'enyi entropy \fer{re} behaves as the Shannon entropy \fer{shan} with respect to the scaling for dilation of the probability density. As usual, for any given density $f(x)$ and positive constant $a$,  we define the \emph{dilation} of $f$ by $a$, as the mass-preserving scaling
 \be\label{scal}
 f(x) \to f_a(x) = {a^n} f\left( {a} x \right).
 \ee
Then, for any $p \ge 0$ it holds
 \be\label{dil}
 \mR_p(f_a) = \mR_p(f) - n \log a.
 \ee
This characteristic differentiates the R\'enyi entropy from other generalizations of the Shannon entropy, which have been introduced later on on the literature. For example, the Tsallis entropy of order $p$ \cite{Tsa}:
 \be\label{tsa}
 \mT_p (f) = \frac 1{1-p}\int_{\pRR^n}\left(  f^p(y) -f(y)  \right)\, dy,
 \ee
which is extensively used by physicists in statistical mechanics \cite{Tsa1}, does not satisfy property \fer{dil}. As we shall see in Section \ref{sec2}, property \fer{dil} is one of the main ingredients to work with the R\'enyi entropy functional and to derive from it inequalities in sharp form. Thus, in our opinion, the definition \fer{re} introduced by R\'enyi represents a very coherent generalization of the Shannon entropy.

The Shannon entropy is naturally coupled to the heat equation (with diffusion coefficient $\kappa$) posed in the whole $\R^n$
  \be\label{heat}
\frac{\partial u(x,t)}{\partial t} =  \kappa \Delta u(x,t),
 \ee
as soon as the initial datum given is assumed to be a probability density.
As recently noticed in \cite{Tos3}, the deep link between the Shannon entropy and the heat equation started to be used as a powerful instrument to obtain mathematical inequalities in sharp form in the years between the late fifties to mid sixties. To our knowledge, the first application of this idea can be found in two
papers  by Linnik \cite{Lin} and Stam \cite{Sta} (cf. also Blachman
\cite{Bla}), published in the same year and concerned with two
apparently disconnected arguments. Stam \cite{Sta} was motivated by
the finding of a rigorous proof of Shannon's entropy power
inequality \cite{Sha}, while Linnik \cite{Lin} used the information
measures of Shannon and Fisher in a proof of the central limit
theorem of probability theory. Also, in the same years, the heat equation has been used in the context of kinetic theory of rarefied gases by McKean \cite{McK} to investigate that large-time behaviour of Kac caricature of a Maxwell gas. There, various monotonicity properties of the derivatives of the Shannon entropy along the solution to the heat equation have been derived.

Likewise, the R\'enyi entropy od order $p$ is strongly coupled to the nonlinear diffusion equation of order $p$ posed in the whole $\R^n$
\begin{equation}
{\partial v(x,t) \over \partial t} =  \kappa \Delta v^p(x,t),
 \label{poro}
\end{equation}
still with the initial datum which is assumed to be a probability density. This maybe not so well-known link has been outlined in some recent papers \cite{CTo,ST}, where various results valid for the Shannon entropy have been shown to hold also for the R\'enyi entropies, and applied to the study of the large-time behavior of the solution to equation \fer{poro}. We aim in this note to highlight this connection.

\section{Nonlinear diffusion equations and self-similarity}
\label{sec2}

To start with, let us recall briefly some essential features of the
nonlinear diffusion equation \fer{poro}. Existence an uniqueness of
the solution of \fer{poro} to the initial value problem posed in the whole space is
well-known \cite{BDV,Vaz}, and we address the interested reader to
these references for details.  The forthcoming analysis will be
restricted to initial data which are probability densities with
finite variance, and it will include both the case $p>1$, usually
known as porous medium equation, and the case $p<1$, the fast diffusion
equation. In dimension $n\ge 1$, the range of exponents which ensure
the boundedness of the second moment of the solution is $p > \bar p
$ with $\bar p = n/(n+2)$, which contains a part of the so-called
fast diffusion range $p < 1$. The particular subinterval of $p$ is
motivated by the existence of a precise solution, found by
Zel'dovich, Kompaneets and Barenblatt in the fifties (briefly
called here Barenblatt solution) \cite{Ba1,Bar,ZK}, which serves
as a model for the asymptotic behavior of a wide class of solutions
with finite second moment. In the case $p>1$ (see \cite{BDV} for $p
<1$) the Barenblatt (also called self-similar or generalized
Gaussian solution)  departing from $x=0$ takes the self-similar form
 \be\label{ba-self}
 M_p(x,t) := \frac 1{t^{n/\mu}} \tilde M_p\left (\frac x{t^{1/\mu}}\right),
 \ee
 where
\[
\mu =2+n(p-1)
\]
and  $\tilde M_p(x)$ is the time-independent function
\begin{equation}
 \tilde M_p(x)=\big(C - \lambda\, |x|^2   \big)^{\frac
1{p-1}}_+ . \label{ba}
 \end{equation}
In \fer{ba} $(s)_+ = \max\{s, 0\}$,
$\lambda = \frac 1{2\mu}\,\frac{p-1}{p}$, and the constant $C$  can be chosen to fix the mass of the source-type Barenblatt
solution equal to one.

The solution to equation \fer{poro} satisfies mass and momentum
conservations, so that
 \be\label{cons}
  \int_{\pRR^n} v(x,t)\, dx =1
\, ; \quad    \int_{\pRR^n} x \, v(x,t)\, dx =  0 \,;  \quad t
 \ge 0.
  \ee
Hence, without loss of generality, one can always assume that
$v_0(x)$ is a probability density of first moment equal to zero. Let
us define by $E(v(t))$  the second moment of the solution:
 \be\label{temp}
 E(v(t)) =  \int_{\pRR^n}|x|^2v(x,t)dx .
 \ee
 Then, $E(v(t))$ increases in time from $E_0 = E(v_0)$, and its
evolution is given by the nonlinear law
 \be\label{ene}
\frac{dE(v(t))}{dt} = 2n \int_{\pRR^n}v^p(x,t)\, dx \ge 0,
 \ee
which is not explicitly integrable unless $p=1$. The second moment
of the solution to equation \fer{poro} has an important role in
connection with the knowledge of the large time behavior of the
solution. Also, in presence of a finite second moment we can
immediately establish a deep connection between equation \fer{poro}
and the R\'enyi entropy of the same order $p$.

Indeed, let us consider the evolution in time of the R\'enyi entropy
of order $p$ along the solution of the nonlinear diffusion equation
\fer{poro}. Integration by parts immediately yields
 \be\label{e-r}
 \frac {d}{dt} \mR _p(v(\cdot,t)) =  \mI_p(v(\cdot,t)), \quad t >0,
 \ee
where, for a given probability density $f(x)$
 \be\label{fis-r}
 \mI_p(f) := \frac 1{\int_{\pRR^n} f^p \, d x}
   \int_{\{f>0\}} \frac{|\nabla f^p(x)|^2}{f(x)} \, d x.
 \ee
When $p \to 1$, identity \fer{e-r} reduces to DeBruijn's identity,
which connects Shannon's entropy functional with the Fisher
information
 \be\label{fis}
\mI(f) = \int_{\{f>0\}} \frac{|\nabla f|^2} f \, dy.
 \ee
via the heat equation \cite{Bla,CT,Sta}. Since $\mI_p(f)
>0$,  identity \fer{e-r} shows that the R\'enyi entropy of the
solution to equation \fer{poro} is increasing in time.

Since the energy scales under the dilation \fer{dil} of $f$
according to
 \be\label{sec}
E(f_a) = \int_{\R^n} |v|^2 f_a(v) \, dv = \frac 1{a^2} E(f),
 \ee
if the probability density has bounded second moment, a dilation
invariant functional is obtained by coupling R\'enyi entropy of $f$
with the logarithm of the second moment of $f$
 \be\label{ent2}
\Lambda_p(f) = \mR_p(f) - \frac n2 \log E(f).
 \ee
Let $v(x,t)$ be a solution to equation \fer{poro}.  If we now
compute the time derivative of $\Lambda_p(v(t))$, we obtain
 \be\label{der1}
\frac d{dt}\Lambda_p(v(t)) = \mI_p(v(t)) - n^2 \frac {\int_{\pRR^n}
v^p(t) \, d x}{E(v(t))},
 \ee
which is a direct consequence of both identities \fer{ene} and
\fer{e-r}.

The right-hand side of \fer{der1} is nonnegative. This can be easily
shown by an argument which is often used in this type of proofs, and
goes back at least to McKean \cite{McK}. One obtains
 \[
0 \le \int_{\{v>0\}}\left( \frac{\nabla v^p(x)}{v(x)} + nx \frac
{\int v^p(x)}{E(v)}\right)^2\, \frac{v(x)}{\int v^p } \,dx  =
 \]
 \[
\mI_p(v) + n^2\frac {\int v^p}{E(v)^2}\int_{\R^n} |x|^2 v(x) \, dx +
2n \frac {\int v^p}{E(v)} \int_{\{v>0\}} x\cdot \nabla v(x)\, dx
=
 \]
 \be\label{pos}
\mI_p(v) + n^2 \frac {\int v^p}{E(v)} -  2n^2\frac {\int v^p}{E(v)}
= \mI_p(v) - n^2 \frac {\int v^p}{E(v)}.
 \ee
Note that equality to zero in \fer{pos} holds if and only if, when
$v(x)>0$
 \[
\frac{\nabla v^p(x)}{v(x)} + n x \frac {\int v^p}{E(v)} = 0.
 \]
This condition can be rewritten as
 \be\label{pos2}
\nabla\left( v^{p-1} + \frac {p-1}{2p} |x|^2\frac{n\int v^p}{E(v)}
\right) = 0
 \ee
which identifies the probability density $v(x)$ as a Barenblatt
density in $\R^n$ (cf. equation \fer{ba}).  Also, \fer{pos} shows that, among all
densities with the same second moment, Fisher information of order
$p$ takes its minimum value in correspondence to a Barenblatt
density.

We proved that the functional \fer{ent2} is monotonically increasing in time along the solution to the nonlinear diffusion. The dilation invariance can now be used to identify the limit value.
The computation of the
limit value uses in a substantial way the scaling invariance
property. Indeed, it is well-known that the solution to equation \fer{poro} converges towards the self-similar Barenblatt solution \fer{ba} in $L_1(\R^n)$ at an explicitly computable rate \cite{BDV,BDGV,CaTo,DD}. By definition, the second moment of the self-similar solution increases in time, and it is infinite as time goes to infinity. However, by dilation invariance, the value of the functional \fer{ent2} in correspondence to a Barenblatt function does not depend on its second moment. In other words, we can scale at each time, without changing the value of the functional, in such a way to fix a certain value of the second moment of the Barenblatt  when time goes to infinity \cite{Tos2,Tos3}. 

The argument we presented is twofold. From one side, it represents a
notable tool to study the large-time behavior of solutions to nonlinear
diffusion equations. From the other side, it allows to find inequalities
by means of solutions to these nonlinear diffusions. Indeed, we proved
that, for any probability density function $f$ with bounded second moment
 \be\label{in12}
\mR_p(f) - \frac n2 \log E(f) \le \mR_p(B_{p,\sigma}) - \frac n2 \log
E(B_{p,\sigma}),
 \ee
where, for $\sigma >0$, we denoted by  $B_{p,\sigma}(x)$ the
Barenblatt density defined in \fer{ba-self}, of second moment equal to $\sigma$. Clearly \fer{in12}
implies that, under a variance constraint, the R\'enyi entropy power
of order $p$ is maximized by a Barenblatt type density.

Inequality \fer{in12} can be rephrased in a slightly different way.
Let $f(x)$ be  a probability density function in $\R^n$, and let
$\mN_p(f)$ denote the entropy power of $f$ associated to the R\'enyi
entropy of order $p$:
 \be\label{epo}
\mN_p(f) = \exp\left\{ \left(\frac 2n + p-1\right)\mR_p(f)\right\}.
 \ee
Then, if  $p> n/(n+2)$,
 \be\label{ep1}
 \frac{\mN_p(f)}{E(f)^{1+ n(p-1)/2}} \le
 \frac{\mN_p(B_{p,\sigma})}{E(B_{p,\sigma})^{1+ n(p-1)/2}}.
 \ee
We note that the definition \fer{epo} of $p$-R\'enyi entropy power,
proposed recently in \cite{ST}, coincides with the classical
definition of Shannon entropy power \cite{Sha}, valid when $p=1$.
This definition requires $p >(n-2)/n$, in which case $2/n + p-1
>0$.
The range of the parameter $p$  for which we can introduce our
notion of R\'enyi entropy power, coincides with the range for which
there is mass conservation for the solution of \fer{poro}
\cite{BDV}. This range includes the cases in which the Barenblatt
has bounded second moment, since $(n-2)/n < n/(n+2)$. We observe
that inequality \fer{ep1} has been derived by a completely different
method in \cite{CHV,LYZ,LYZ2}.

\section{The concavity of R\'enyi entropy power}
\label{sec3}

The Shannon entropy power of a probability density,
 \be\label{sep}
 \mN(f) = \exp\left\{ \frac 2n \mR(f)\right\},
 \ee
where $\mR(f)$, the Shannon entropy, has been defined in \fer{shan},
is one of the most important concepts of information theory
\cite{CT}. Starting on it, Shannon derived the celebrated entropy
power inequality \cite{Bla,Sta}, which reads
 \be\label{entr}
\mN(f*g)\ge \mN(f) + \mN(g).
 \ee
In \fer{entr}, $f$ and $g$ are two probability density functions,
and $f*g$ denotes convolution. Equality in \fer{entr} holds if and
only if both $f$ and $g$ are Gaussian densities.

Also, other properties of Shannon entropy power $\mN(f)$ have been
discovered. In particular, the \emph{concavity of entropy
power}, which asserts that, if $u(x,t)$ is a solution to the heat
equation \fer{heat}, corresponding to an initial datum $u_0(x)$ that
is a probability density, then
 \be\label{conc}
\frac{d^2}{dt^2}\mR(u(\cdot, t)) \le 0.
 \ee
Moreover, equality in \fer{conc} holds if and only if $u(x,t)$
coincides with the Gaussian density of variance $t$, namely the
self-similar solution to the heat equation.  Inequality \fer{conc}
is due to Costa \cite{Cos}. More recently, a short and clean proof
of \fer{conc} has been obtained by Villani \cite{Vil}, by means of an argument 
introduced by McKean \cite{McK} in his paper on Kac caricature of a Maxwell gas. Recently \cite{Tos2}, we
investigated various not so well-known consequences of the concavity property \fer{conc}. Among others, concavity implies the so-called isoperimetric inequality for entropies
\cite{DCT}. For any probability density $f(x)$ in $\R^n$, this
inequality asserts that
  \be\label{b51}
\mN(f) \, \mI(f) \ge   \mN( M_\sigma )\, \mI(M_\sigma),
 \ee
where $\mI(f)$ is the Fisher information of $f$ defined in \fer{fis}
and, for any constant $\sigma
>0$, $M_\sigma$ denotes the gaussian density in $\R^n$ of mean zero,
and variance $\sigma$, that is
 \be\label{Gauss}
M_\sigma(x) = \frac 1{(2\pi\sigma)^{n/2}} \exp\left\{-
\frac{|x|^2}{2\sigma}\right\}.
 \ee
As for \fer{in12}, inequality \fer{b5} implies that, the product of
the Shannon entropy power and the Fisher information is minimized by
a Gaussian density.

The physical idea behind the concavity of the Shannon entropy power
is clear. If we evaluate the entropy power in correspondence to a
Gaussian density like \fer{Gauss}, we obtain
 \[
\mN( M_\sigma ) = 2 \pi \sigma e.
 \]
Hence, the entropy power of the self similar solution to the heat
equation, namely a Gaussian density of variance $2t$, is a linear
function of time, and its second derivative (with respect to time)
is equal to zero. This property is restricted to Gaussian densities.
Any other solution to the heat equation, different from the
self-similar one, is such that its entropy power is concave.

Having in mind to extend the concavity property to the R\'enyi entropy
power, and making use of the result of Section \ref{sec2}, in which
we  established a connection of the R\'enyi entropy with the solution
of the nonlinear diffusion equation, the starting point for the
proof of such a property would be a definition of R\'enyi entropy
power (of order $p$) which is consistent with the fact that, when
evaluated in correspondence to the Barenblatt self-similar solution
\fer{ba-self} to the nonlinear diffusion of order $p$, the value of
the R\'enyi entropy power is linear with respect to $t$. It is a
simple exercise to verify that, owing to definition \fer{epo}, this
is true, since
 \be\label{lin2}
\mN_p(M_p(t)) = \mN_p(\tilde M_p)\cdot t.
 \ee
In \cite{ST}, starting from definition \fer{epo}, we proved that the
R\'enyi entropy power of order $p$ has the concavity property when
evaluated along the solution to the nonlinear diffusion \fer{poro}.
The precise result is the following:

\begin{theorem} ({\bf\cite{ST}})
  Let  $p>(n-2)/n$ and let $u(\cdot,t)$ be probability densities in
  $\R^n$ solving \fer{poro} for $t>0$.
  Then the  $p$-th R\'enyi entropy power
  defined by \fer{epo} satisfies
   \be\label{conc-p}
   \frac{d^2}{d
    t^2}\mN_p(v(\cdot,t)) \le 0
     \ee
\end{theorem}
Like in the Shannon's case, inequality \fer{conc-p} lieds to sharp
isoperimetric inequalities. The (isoperimetric) inequality for the
$p$-th R\'enyi entropy is contained into the following
\begin{theorem}({\bf\cite{ST}})
  If $p > n/(n+2)$
  every smooth, strictly positive and rapidly decaying probability
  density $f$
  satisfies
  \be\label{b5}
  \mN_p(f) \, \mI_p(f) \ge \mN_p(\tilde
  M_p) \, \mI_p(\tilde M_p) =
    \gamma_{n,p}.
  \ee
\end{theorem}
We remark that $\mI_p(f)$ is the generalized Fisher information
defined in \fer{fis-r}. Once again, it is immediate to show that the product in \fer{b5} is invariant under dilation, which allows to reckon explicitly the value of the constant by using the same argument of Section \ref{sec2}.  If $p>1$ the value of the
constant $\gamma_{n,p}$ is
 \be\label{pg1}
 \gamma_{n,p} =  n\pi\frac{2p}{p-1}\left(
\frac{\Gamma\left(\frac{p+1}p\right)}{\Gamma\left(\frac n2
+\frac{p+1}p \right)}\right)^{2/n}\left(\frac{(n+2)p -n}{2p}
\right)^{\frac{2+ n(p-1)}{n(p-1)}}.
 \ee
In the remaining set of the parameter $p$, that is if $n/(n+2) < p
<1$,
 \be\label{pl1}
\gamma_{n,p} =  n\pi\frac{2p}{1-p}\left(\frac{\Gamma\left( \frac
1{1-p} -\frac n2 \right)}{\Gamma\left(\frac 1{1-p}\right)}
\right)^{2/n}\left(\frac{(n+2)p -n}{2p} \right)^{\frac{2+
n(p-1)}{n(p-1)}}.
 \ee
Inequality \fer{b5} can be rewritten in a form more suitable to
functional analysis. Let $f(x)$ be a probability density in $\R^n$.
Then, if $p > n/(n+2)$
 \be\label{gn}
\int_{\R^n} \frac{|\nabla f^p(x)|^2}{f(x)} \, d x \ge \gamma_{n,p}
\left( \int_{\R^n} f^p(x) \, d x \right)^{\frac{ 2+ 2n
(p-1)}{n(p-1)}}.
 \ee
If $n >2$, the case $p= (n-1)/n$ is distinguished from the others,
since it leads to
 \[
\frac{2+ 2n (p-1)}{n(p-1)}= 0, \quad  \nu=\frac 1n,
 \]
 and
\begin{displaymath}
  \mN_{1-1/n}(f)=\int_{\R^n}f^{1-1/n}(x)\,d x.
\end{displaymath}
In this case the concavity of $\mN_{1-1/n}$ along \fer{poro} has
been already known and has a nice geometric interpretation in terms
of transport distances, see \cite{Otto01}.

Note that the restriction $n>2$ implies $(n-1)/n > n/(n+2)$. Hence,
for $p= (n-1)/n$ we obtain that the probability density $f$
satisfies the inequality
 \be\label{sob12}
\int_{\R^n} \frac{|\nabla f^{(n-1)/n}(x)|^2}{f(x)} \, d x \ge
\gamma_{n,(n-1)/n}.
 \ee
The substitution $f= g^{2^*}$, where $2^*=2n/(n-2)$, yields
 \[
 \int_{\R^n} \frac{|\nabla f^{(n-1)/n}(x)|^2}{f(x)} \, d x =
 \left(\frac{2n-2}{n-2}\right)^2\int_{\R^n} |\nabla g(x)|^2 \,
 d x.
 \]
Therefore, for any given function $g \ge 0$ such that $g(x)^{2^*}$
is a probability density in $\R^n$, with $n>2$, we obtain the
inequality
 \be\label{sob14}
\int_{\R^n} |\nabla g(x)|^2 \,
 d x \ge
\left(\frac {n-2}{2n-2}\right)^2\,\gamma_{n,(n-1)/n}.
 \ee
Since
 \[
\gamma_{n,(n-1)/n} = n\pi\,
\frac{2^2(n-1)^2}{n-2}\left(\frac{\Gamma\left(
 n/2 \right)}{\Gamma\left(n \right)} \right)^{2/n},
 \]
a simple scaling argument finally shows that, if $g(x)^{2^*}$ has a
mass different from $1$, $g$ satisfies the \emph{Sobolev} inequality
\cite{Aub}, \cite{Tal}
 \be\label{sob15}
 \int_{\R^n} |\nabla g(x)|^2 \,
 d x
\ge \mathcal{S}_n \left(\int_{\R^n} g(x)^{2^* 
  }\,d x \right)^{2/2^*},
 \ee
where
 \[
\mathcal{S}_n  = n(n-2) \pi \, \left(\frac{\Gamma\left(
 n/2 \right)}{\Gamma\left(n \right)} \right)^{2/n}
 \]
is the sharp Sobolev constant. Hence, Sobolev inequality with the
sharp constant is a consequence of the concavity of  R\'enyi entropy
power of parameter $ p = (n-1)/n$, when $n >2$.

In all the other cases, the concavity of R\'enyi entropy power leads
to Gagliardo-Nirenberg type inequalities with sharp constants, like
the ones recently studied by Del Pino and Dolbeault \cite{DD}, and
Cordero-Erausquin, Nazaret, and Villani, \cite{CNV} with different
methods.

\section{Conclusions}

In this note we presented recent results which highlight the links between the R\'enyi entropies and the nonlinear diffusion equations. This connection is at the basis of an information-theoretic proof of various functional inequalities, which, at difference with previous approaches, have the merit to be intuitive. The main example is furnished by the result of concavity presented in Section \ref{sec3}, where the classical sharp Sobolev inequality is derived as a particular case of the concavity property of the R\'enyi entropy of order $1-1/n$. Also, the inequalities presented here can be fruitfully employed to deeply investigate the large-time behavior of the solution to nonlinear diffusion equations \cite{CTo}. Other problems remain to be investigated.  Mainly, it would be important to understand if the Shannon entropy power \fer{entr}, which deals with convolutions, admits a version in terms of R\'enyi entropy power, whose proof is based on this connection with nonlinear diffusion equations.




\end{document}